# Measurement theory of a density profile of small spheres on a cylindrical surface: Conversion of force curve measured with surface force apparatus into pressure on its surface element

Kota Hashimoto[a] and Ken-ichi Amano[a]

[a]*Department of Energy and Hydrocarbon Chemistry, Graduate School of Engineering, Kyoto University, Kyoto* 615-8510, *Japan*.

E-mail: kotah314@yandex.com

**ABSTRACT**
Recently, in an ensemble of small spheres, we proposed a method that converts the force between two large spheres into the pressure on the large sphere's surface element. Using it, the density distribution of the small spheres around the large sphere can be obtained experimentally. In a similar manner, in this letter, we propose a transform theory for surface force apparatus, which transforms the force acting on the cylinder into the density distribution of the small spheres on the cylindrical surface. The transform theory we derived is briefly explained in this letter.

**MAIN TEXT**

Structure of liquid-solid interface is an important knowledge for elucidations of mechanisms of several interactions at the interface. To profoundly understand reactions at the interface, solvation structure on a solid or a membrane should be known. Furthermore, a density distribution of colloid particles on the solid or the membrane is also an important thing, because it is much related to development of a study of soft matter. The density distribution of the colloid particles attracts attention in fundamental studies of biophysics, tribology, cosmetology, and so on. In this short letter, we explain how to obtain the solvation structure and the density distribution of colloid particles from an experimental datum. In the explanation, solvent molecules and colloid particles are collectively treated as small spheres, and surface of the solid and the membrane are collectively modeled as that of a cylinder. We show that a force curve between the two cylinders can be transformed into the density distribution of small spheres on the cylinder through our theory derived here, where the two cylinders are perpendicularly crossed (see Figs. 1-3). Since the force curve can be measured by surface force apparatus (SFA) [1-3], the density distribution can be estimated from the experimental datum. The transform theory is derived on the bases of statistical mechanics of simple liquids [4] and recently proffered transform theories [5-10]. The derivation process of the transform theory for the crossed cylinders is explained below.

Firstly, we explain the theoretical model system. In the model system, there are two perpendicularly crossed cylinders 1 and 2 (see Figs. 1-3), which are immersed in an ensemble of small spheres. Cylinder 1's central axis is identical to $y$-axis, and cylinder 2's central axis is identical to $x$-axis. Radius of the cylinder is expressed as $r_C$, and its rod length is much longer than $r_C$. Radius of the small sphere is expressed as $r_S$, and bulk number density of the small spheres is $\rho_0$. We define that $r = r_C + r_S$, where $r$ is radius of the excluded volume of the cylinder (a space which the center of the small sphere cannot enter is excluded volume). In Fig. 3, $s$ represents separation between the nearest surfaces of the cylinders, and $l$ is length between the facing surface elements of the cylinders. Meaning of $\theta$ (radian) is shown in the figure. In the model system, two-body potential between the two small spheres is arbitrary, and that between the two cylinders is also arbitrary. However, two-body potential between the cylinder and the small sphere is rigid. Under the model system, we focus on an oscillatory force on the cylinder 1 and derive the transform theory. The oscillatory force indicates a solvation force or colloid-induced force, which does not contain two-body force between the two cylinders. Here, we briefly mention the oscillatory force as force. If $z$ direction force acting on the cylinder 1 is expressed by a summation of forces between the face-to-face surface elements, it can be expressed as

$$f(s) = \int_{C1} P(l) dA_z, \qquad (1)$$

where $P$ is $z$ direction pressure on the surface element of the cylinder 1. $dA_z$ is a minimal efficient area of the surface element of the cylinder 1, which is normal to $z$-axis. C1 means the integration is a surface integral which is performed on surface of the excluded volume of the cylinder 1. We notify that $P$ is equal to a sum of pressures on the upper and under sides. In the present case, Eq. (1) is rewritten as

$$f(s) = 4 \int_0^{\frac{\pi}{2}} \int_0^r P\left(s + 2r - r\cos\theta - \sqrt{r^2 - y^2}\right) \cos\theta \, r \, dy \, d\theta. \tag{2}$$

To change the integral equation above into a solvable form, we introduce following three variable transformations in turn:

$$y = r\sqrt{1 - v^2}, \tag{3}$$

$$\cos\theta = (s - w)/r, \tag{4}$$

$$a = w + 2r - rv. \tag{5}$$

Then, Eq.(2) is changed to

$$f(s) = 4 \int_{s-r}^{s} \int_{w+r}^{w+2r} P(a) \frac{w + 2r - a}{\sqrt{r^2 - (w + 2r - a)^2}} da \frac{s - w}{\sqrt{r^2 - (s - w)^2}} dw. \tag{6}$$

In this form, the integral equation can be solved by using a matrix operation [9,10]. To obtain $P$, Eq. (6) is written in the matrix form below:

$$\boldsymbol{f} = 4\boldsymbol{HMP}, \tag{7}$$

where $\boldsymbol{f}$ and $\boldsymbol{P}$ corresponds to $f(s)$ and $P(a)$, respectively, and $\boldsymbol{H}$ and $\boldsymbol{M}$ are square matrixes. If cells of $\boldsymbol{H}$ and $\boldsymbol{M}$ are represented as $H_{ij}$ and $M_{ij}$, respectively, in the integral ranges, they can be written as

$$H_{ij} = \frac{s_i - w_j}{\sqrt{r^2 - (s_i - w_j)^2}} \Delta w, \tag{8}$$

$$M_{ij} = \frac{w_i + 2r - a_j}{\sqrt{r^2 - (w_i + 2r - a_j)^2}} \Delta a, \tag{9}$$

where $\Delta w$ and $\Delta a$ are sufficiently small values. In the external ranges, $H_{ij}$ and $M_{ij}$ are both zero. However, at the integral boundaries, $H_{ij}$ and $M_{ij}$ become infinity. Hence, we alternatively express $H_{ij}$ and $M_{ij}$ in the integral ranges as follows:

$$H_{ij} = \int_{w_j}^{w_{j+1}} \frac{s_i - w}{\sqrt{r^2 - (s_i - w)^2}} dw, \tag{10}$$

$$M_{ij} = \int_{a_j}^{a_{j+1}} \frac{w_i + 2r - a}{\sqrt{r^2 - (w_i + 2r - a)^2}} da. \tag{11}$$

Using these expressions, we can avoid the infinity, because they have analytical solutions within certain values. In addition, these expressions are *almost completely exact*, when values of "$w_{j+1} - w_j$" and "$a_{j+1} - a_j$" are very small. This idea was originated from a following approximation (which is written in a general form):

$$\int_{c_j}^{c_{j+1}} A(x)B(x)dx \cong A(c_j) \int_{c_j}^{c_{j+1}} B(x)dx. \tag{12}$$

We notify that there is no need to care about deviation which comes from the approximation, when the integral range is sufficiently short and absolute value of "$A(c_{j+1}) - A(c_j)$" is sufficiently small. Consequently, we can calculate $\boldsymbol{P}$ from $\boldsymbol{f}$ by utilizing an inverse matrix of $\boldsymbol{HM}$. The change from $\boldsymbol{f}$ to $\boldsymbol{P}$ is FPSE conversion [9,10]. By the way, the change can be performed also by Derjaguin approximation [11-13]. Comparison between the present FPSE conversion and Derjaguin approximation will be conducted in our future study.

Density distribution of the small spheres around the cylinder can be calculated by applying the transform theories for two-flat rigid surfaces [6], which we call rigid wall (RW) theory. By using RW theory, the normalized density distribution being $g$ is obtained as follows:

$$g(r + l) = \frac{P(l)}{k_B T \rho_0 g_c} + 1, \tag{13}$$

where

$$g_c = \frac{1 + \sqrt{1 + 4P(0)/(k_B T \rho_0)}}{2}. \tag{14}$$

Here, $k_B$ and $T$ are the Boltzmann constant and absolute temperature, respectively. gc represents the normalized number density at a contact point between the cylinder and the small sphere. Eventually, it can be said that the transform theory is a hybrid of FPSE conversion for the cylinders and RW theory.

We verified the transform theory explained here by using a computer. The first step verification tests were conducted in a rigid system and in a Lennard-Jones fluid. Firstly, the normalized density distribution of the solvent around the cylinder ($g_B$) is prepared by using three-dimensional Ornstein-Zernike equation coupled with hypernetted-chain closure (3D-OZ-HNC) [4]. Here, $g_B$ is a benchmark for this verification test. Next, the input data (force curve between the two cylinders) is calculated also by using 3D-OZ-HNC. The input data $f$ is converted to the pressure $P$, and $P$ is transformed into the density distribution $g$ of the output. We have found that the output $g$ is similar to $g_B$ (not shown here).

In summary, we have proposed the transform theory for calculating the density distribution of the small spheres around the cylinder from the force curve obtained by SFA with the cylinders. We have written the derivation process of the transform theory in detail. It has been found that the integral equation connecting the force and the pressure can be solved by combining the variable transformations, avoidance of the infinity, and the matrix operation. In the near future, we will show results of the verification tests of the transform theory.

ACKNOWLEDGEMENTS
We appreciate T. Sakka, N. Nishi, T. Fukuma, K. Fukami, and H. Matsubara for the useful advice and discussions. This work was supported by "Grant-in-Aid for Young Scientists (B) from Japan Society for the Promotion of Science (15K21100)".

**FIGURES**

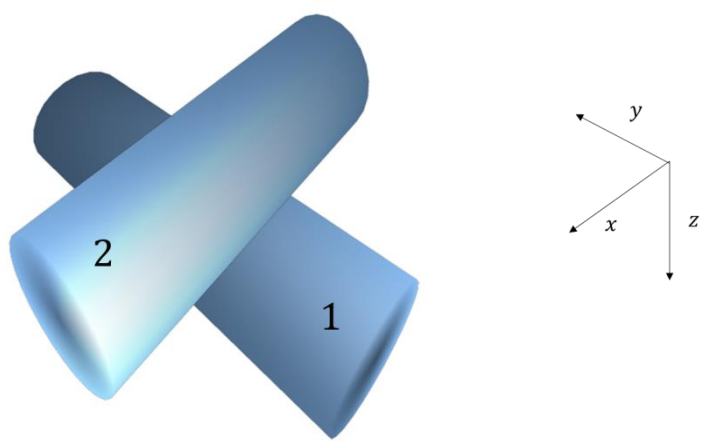

Fig.1 Illustration of the perpendicularly crossed cylinders.
The cylinders 1 and 2 are immersed in an ensemble of small spheres.

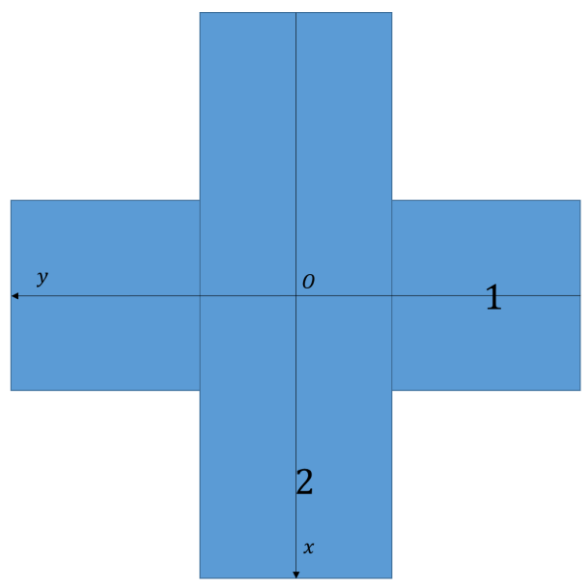

Fig.2 Illustration of the two cylinders viewed from $z$-axis. $O$ indicates (x,y) = (0,0).

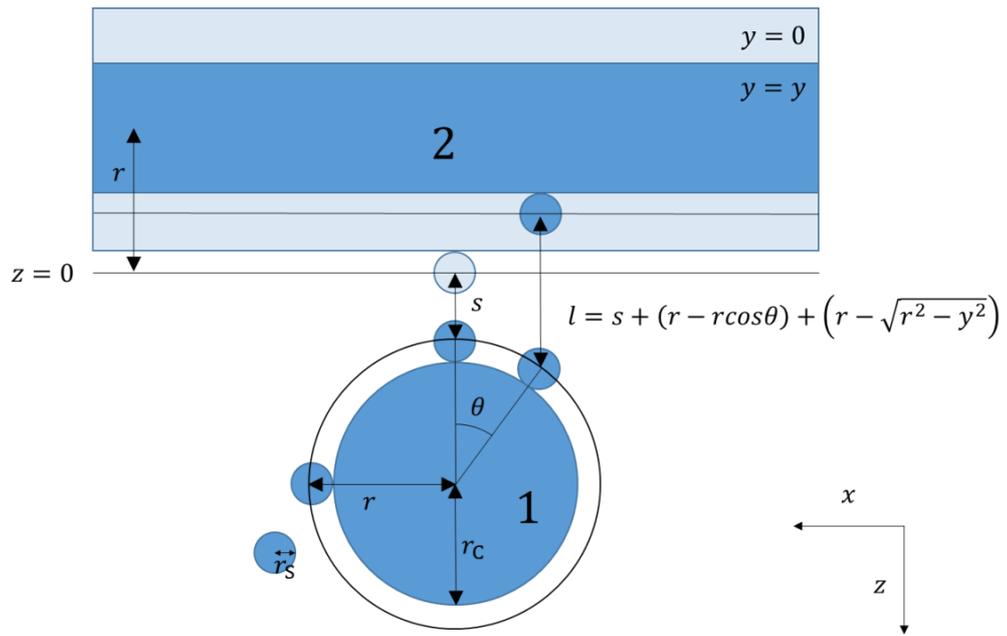

Fig.3 Illustration of the two cylinders viewed from *y*-axis.
Cross section of cylinder 2 at *y* = 0 is colored by faint blue, and the cross section of is at *y* = *y'* is colored dark blue (*y'* is an arbitrary value in the range from −*r* to +*r*).